\documentclass[
]{scrarticle}
\usepackage{amsmath,amssymb} 
\usepackage{graphicx}        
\usepackage{xcolor}          
\usepackage{authblk}         

\newcommand{\Hctrl}{H_\mathrm{ctrl}}

\newcommand{\COMMENT}[1]{}

\date{}

\begin{document}

\title{Machine Learning-Aided Optimal Control of a Qubit Subjected to External Noise}

\author[1]{Riccardo Cantone}
\author[1]{Shreyasi Mukherjee}
\author[1,2]{Luigi Giannelli}
\author[1,2,3]{Elisabetta Paladino}
\author[1,2,3]{Giuseppe A. Falci}

\affil[1]{Dipartimento di Fisica e Astronomia ``Ettore Majorana'', Universit\`a di Catania, Via S. Sofia 64, 95123 Catania, Italy}

\affil[2]{Istituto Nazionale di Fisica Nucleare, Sezione di Catania, 95123, Catania, Italy}

\affil[3]{CNR-IMM, Via S. Sofia 64, 95123, Catania, Italy}

\maketitle
\vspace{-1.5cm}
\begin{abstract}

    We apply a machine-learning-enhanced greybox framework to a quantum optimal control protocol for open quantum systems. Combining a whitebox physical model with a neural-network blackbox trained on synthetic data, the method captures non-Markovian noise effects and achieves gate fidelities above 90\% under Random Telegraph and Ornstein-Uhlenbeck noise. Critical issues of the approach are discussed. 
\end{abstract}

\section{Introduction}

This work showcases an attention-based machine-learning-enhanced \textit{greybox} framework for quantum optimal control, designed to improve the manipulation of open quantum systems subject to complex noise~\cite{youssri1,attention}. Quantum control~\cite{giannelli} is essential for quantum technologies such as computation, communication, and sensing, yet achieving robust control remains challenging when the system interacts with non-Markovian environments that are hard or even impossible to characterize.

The proposed greybox model combines:
\begin{itemize}
  \item a \emph{whitebox} component that captures the analytically tractable portion of the system dynamics using known physical principles;
  \item a \emph{blackbox} component implemented via neural networks, trained to learn the unmodelled effects of the environment from data.
\end{itemize}

The model is tested using synthetic data by applying it to a qubit undergoing pure dephasing. We consider two types of environmental noise: Random Telegraph Noise (RTN) and Ornstein-Uhlenbeck (OU) noise~\cite{paladino}. Synthetic datasets are generated through simulations of the stochastic Schr\"odinger dynamics, with input-output pairs linking control parameters (i.e., external drive amplitudes) to gate fidelities. The gate fidelities are related to a universal set of single qubit gates. The blackbox component is trained in a supervised fashion to infer parameters encoding the impact of the environment. Combining the whitebox and blackbox components enables accurate emulation of the evolution of the principal system across a wide range of coupling strengths with the environment.

After training, a gradient-based optimal control method is employed to design control pulses for implementing a universal set of single-qubit gates. Gate fidelities above $99\%$ are achieved at low values of noise coupling, and remain above $90\%$ even at the strongest noise regimes considered.

  \section{\label{sec:system}System and Model}
  We consider a single qubit subject to classical dephasing noise along the $z$-axis
  and driven by external control fields. In the interaction picture, the dynamics
  is described by the time-dependent Hamiltonian 
  \begin{equation*}
    H(t) = \Hctrl(t) + g\beta(t)\,\sigma_{z}, 
  \end{equation*}
  where $g$ is the coupling strength between the qubit and the noise, and
  $\beta( t)$ is a classical stochastic process modeling dephasing noise
  \cite{auza}. Specifically, in this work we consider $\beta(t)$ to be either a
  Random Telegraph Noise (RTN) process or an Ornstein-Uhlenbeck (OU) process.

  The control Hamiltonian $\Hctrl(t)$ implements a drive along the $x$ and $y$-axes
  \begin{equation*}
    \label{eq:cntrlham}\Hctrl(t) = f_{x}(t)\sigma_{x}+ f_{y}(t)\sigma_{y},
  \end{equation*}
  each control field $f_{\alpha}(t)$, with $\alpha \in \{x, y\}$, consisting
  of $5$ Gaussian-shaped pulses.

Two different stochastic processes were considered and compared, namely an RTN process and an OU process. They are characterised by their power spectrum $S(\omega)$, which is the Fourier transform of the two-point correlation function $\langle \beta(t) \beta(t) \rangle$, and in both cases has a Lorentzian shape~\cite{paladino, falci4} 
\begin{equation}
    S(\omega)\approx \frac{4\gamma}{4\gamma^2+\omega^2}
\end{equation}
where $\gamma$ is the switching rate for the RTN process and $1/\gamma$ is the correlation time of the OU process. The two stochastic processes differ since the latter is Gaussian, while the former is not, and it is known that this has an impact on dynamic protocols of protection against noise, such as spin-echo~\cite{paladino,bergli_decoherence_2009}.

\section{The Machine Learning Model}

The proposed greybox model integrates analytical knowledge of the quantum system with a transformer-based neural network. This hybrid architecture includes two components:

\begin{itemize}
\item A \emph{whitebox} part, which enforces the known unitary dynamics of the driven qubit and the associated measurement process;
\item A \emph{blackbox} neural network, trained to model the influence of the environment on the system's evolution.
\end{itemize}

\paragraph{Inputs and Outputs}

The model takes as input the amplitudes of five Gaussian control pulses applied along each of the $x$ and $y$ axes, for a total of ten real parameters. The pulse widths and positions are fixed. The output consists of six gate fidelities, each associated with a different target from a universal set of single-qubit gates.

\paragraph{Model Architecture}

The blackbox core is a lightweight transformer encoder. It processes the input pulse parameters and predicts a set of noise-related parameters that are fed into whitebox layers implementing:

\begin{itemize}
\item Hamiltonian construction and time evolution based on discretized control fields;
\item Expectation value calculation over a tomographically complete set of initial states;
\item Process matrix reconstruction and fidelity estimation.
\end{itemize}

In addition, the model includes specialized output heads that refine the predicted expectation values before computing the final fidelities.

\paragraph{Training Strategy}

Only the blackbox layers contain trainable parameters. The network is trained using the Adam optimizer to minimize the mean squared error across the six predicted fidelities. Training is supervised and based on synthetic data generated by simulating noisy quantum dynamics. Whitebox constraints ensure physically consistent predictions throughout.

\begin{figure}[h]
    \centering
    \includegraphics[width=1\textwidth, trim=0.0cm 0cm 0.0cm 0cm, clip]{
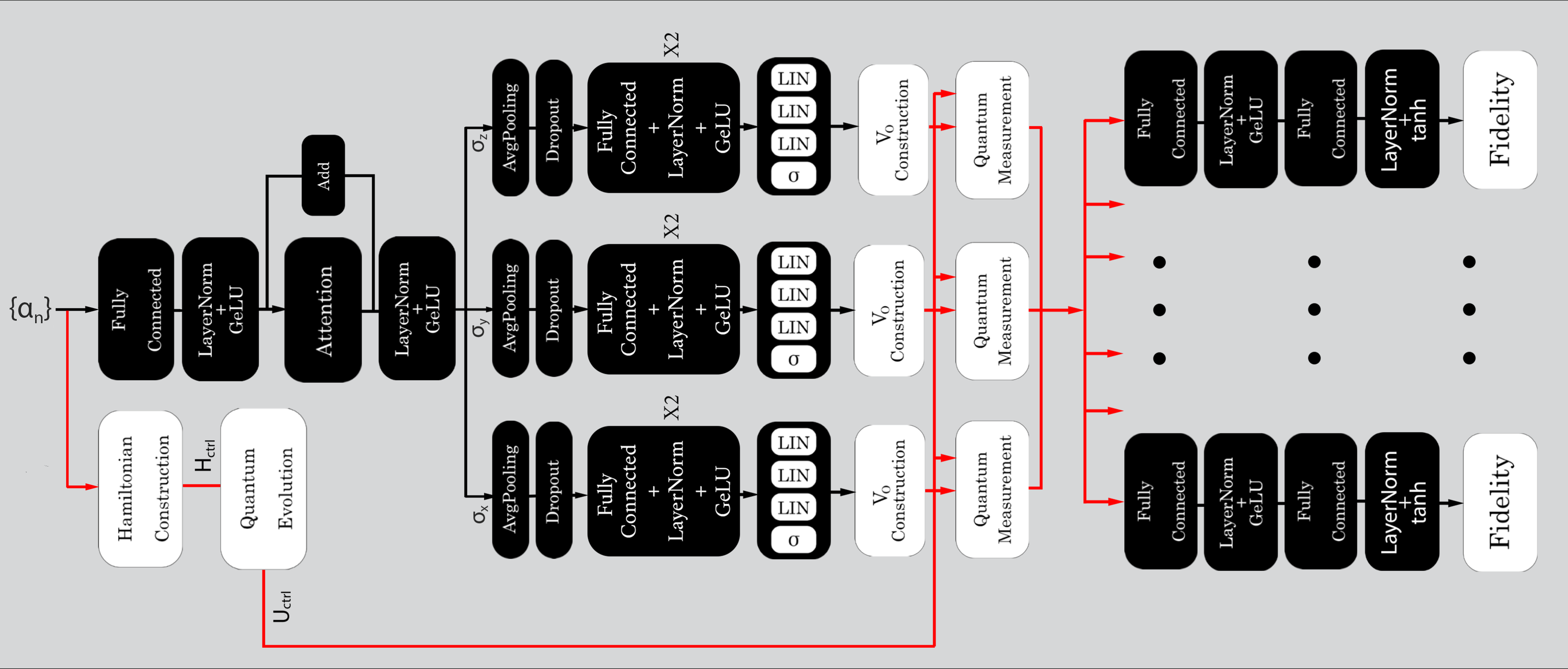
    }
    \caption{The transformer-based graybox architecture we used to model
    Markovian and non-Markovian open-system qubit dynamics.}
    \label{NN}
  \end{figure}

A schematic overview of the architecture is shown in Fig.~\ref{NN}.

\section{Results and open problems}

Separate models were trained across varying values of the coupling strength $g$. This provides insight on the effectiveness of the graybox approach as a function of the Markovianity of the quantum map describing time evolution under the effect of noise, which in this case is parametrized by the ratio $g/\gamma$~\cite{paladino_decoherence_2003,paladino}.

\paragraph{RTN Case.} The model showed low training and test MSE across all gates, with prediction errors increasing with $g$ but remaining in the $10^{-2}$--$10^{-3}$ range, indicating robust generalisation. As an emulator in the optimal control pipeline, it enabled the design of control pulses achieving fidelities above 99\% for the lowest $g$ and above 90\% for the highest, with minor gate-dependent variations.
\paragraph{OU Case.} The model exhibited similar performance, with low and stable MSE values across all $g$, confirming robustness to different noise types. Optimal control results mirrored those of the RTN case, with fidelities exceeding 99\% at low $g$ and remaining above 90\% even at stronger coupling. While fidelity declines under higher noise, the model continues to support effective pulse design; future improvements may benefit from larger datasets or more advanced strategies.\\

Our result validates the graybox approach, showing that the optimization framework we have chosen is very effective in suppressing effects of low-frequency noise ($g/\gamma > 1$), but less effective for noise yielding  Markovian maps  ($g/\gamma < 1$). Apparently, Gaussianity does not have an impact in this case, but we expect that the picture may change when considering $1/f$ noise~\cite{paladino_decoherence_2002,paladino,falci4,otterpohl_quantum_2025} resulting from a set of stochastic processes with different $\gamma$.

A natural development of this work is applying the method to two-qubit gates, addressing the effect of both time- and space-correlated noise~\cite{darrigo_effects_2008,darrigo_open-loop_2024,fasone_detection_2025}. Two major issues to be investigated are the scalability of the approach to larger quantum architectures and the ability to reproduce asymptotic results known from the theory of dynamical decoupling~\cite{falci_dynamical_2004,paladino,darrigo_open-loop_2024}.

\section*{Acknowledgements}

This work was partially funded by the PNRR MUR project PE0000023 National Quantum Science and Technology Institute (NQSTI), the Centro Nazionale di
Ricerca ICSC in High Performance Computing, Big Data e Quantum Computing
(ICSC), and the Gruppo Nazionale per il Calcolo Scientifico (GNCS) of the Istituto Nazionale di Alta Matematica “Francesco Severi”.

\bibliographystyle{unsrt}
\bibliography{bibliography}

\end{document}